\documentclass[a4paper,aip,preprint]{revtex4-2}
\usepackage{graphicx}
\usepackage{dcolumn}
\usepackage{amsmath}
\usepackage{fancyhdr}
\usepackage{float}
\usepackage{dblfloatfix}
\usepackage{cancel}
\usepackage{amsbsy}
\usepackage{color}
\usepackage{natbib}
\usepackage{appendix}
\usepackage[export]{adjustbox}
\usepackage[caption =false]{subfig}
\setcitestyle{round}
\begin{document}

\title{Thermal behavior of small magnets}

\author{Lukas Herron}
\affiliation{Department of Physics, University of Florida, Gainesville, Florida 
32611, USA}%
\author{Purushottam Dixit}
\affiliation{Department of Physics, University of Florida, Gainesville, Florida 
32611, USA}
\affiliation{UF Genetics Institute, University of Florida, Gainesville, Florida 
32611, USA}%
\date{\today}

\begin{abstract}
While the canonical ensemble has been tremendously successful in capturing statistical properties of large systems, deviations from canonical behavior exhibited by small systems  are not well understood. Here, using a two dimensional small Ising magnet embedded inside a larger heat bath, we characterize the failures of the canonical ensemble when describing small systems. We find significant deviations from the canonical behavior for small systems near and below the critical point. Notably, the agreement with the canonical ensemble is driven not by the system size but by the decoupling between the system and its surrounding. A superstatistical framework wherein we allow the temperature of the small magnet to vary is able to capture the statistics of the small magnet with significantly higher accuracy than the Gibbs-Boltzmann distribution. We discuss implications for experiments and future directions. 
\end{abstract}
\maketitle

{\bf Introduction:} Recent advancements in magnetic resonance imaging have enabled individual spins in magnetic structures and atomic sized magnets to be observed \citep{Maurer2010}. Given their significant importance in the miniaturization of electronic devices \citep{Carlton2008}, there is great interest in controlling the spin configurations of these small magnets using external control variables such as temperature and magnetic fields~\citep{rotskoff2017geometric}.

In general, in order to manipulate the properties of a thermally fluctuating system, we first need to understand its statistics; the equilibrium probability distribution of its states as the system traverses through the phase space. A natural choice to model the thermal statistics of small magnets (the `system') embedded in and exchanging energy with a larger magnet (the `surroundings') is the Gibbs-Boltzmann distribution. Here, the probability of observing any particular configuration of the system is given by
\begin{eqnarray}
p({\bf x}) = \frac{1}{Z(\beta)} \exp \left ( -\beta \varepsilon({\bf x}) \right ). \label{eq:canon}
\end{eqnarray}
In Eq.~\ref{eq:canon}, ${\bf x}$ is the configuration of the system (for example, the collective state of all spins of a magnet), $\varepsilon({\bf x})$ is the Hamiltonian, and $\beta$ is the inverse temperature. The Gibbs-Boltzmann distribution posits that the complex interactions between the system and its surroundings can be captured effectively by a single parameter; the inverse temperature $\beta$. Remarkably, this simple description accurately predicts the thermal properties of a wide range of systems. The Gibbs-Boltzmann distribution is derived with the help of two assumptions~\citep{chandler1987introduction}. First, we assume that  the system under consideration is macroscopic. This assumption allows us to  neglect the `boundary' interactions between the system and the surroundings. That is, we can write: 
\begin{eqnarray}
\varepsilon_{\rm tot}({\bf x}, {\bf y})  &=& \varepsilon({\bf x})  + \varepsilon_{\rm bath}({\bf y})  + \varepsilon_{\rm int}({\bf x}, {\bf y}) \nonumber \\
&\approx& \varepsilon({\bf x}) + \varepsilon_{\rm bath}({\bf y}).
\end{eqnarray}
The second assumption states that the system is much smaller than the surroundings, that is,  $\varepsilon_{\rm tot}({\bf x}, {\bf y}) \gg \varepsilon({\bf x})$. With these two key assumptions, we can conclude that when the total energy $\varepsilon_{\rm tot}$ is kept constant, the number of ways in which the system can be in configuration ${\bf x}$ is exactly equal to the number of ways in which the bath will have energy $\varepsilon_{\rm tot} - \varepsilon({\bf x})$. This number is directly proportional to the probability of observing the system in configuration ${\bf x}$ and is given by
\begin{eqnarray}
p({\bf x}) &\propto& \Omega \left ( \varepsilon_{\rm tot} - \varepsilon({\bf x})  \right ) = \exp \left ( S_{\rm bath} \left (  \varepsilon_{\rm tot} - \varepsilon({\bf x}) \right )\right )  \nonumber \\
&\approx& \exp \left ( S_{\rm bath} \left (  \varepsilon_{\rm tot}  \right ) - \beta \varepsilon({\bf x})\right )  \propto \exp \left ( -\beta \varepsilon({\bf x}) \right ). \label{eq:canderive}
\end{eqnarray}
Here, $S_{\rm bath}$ is the microcanonical entropy of the bath and a Taylor series approximation is invoked to approximate    $S_{\rm bath}(\varepsilon_{\rm tot} - \varepsilon({\bf x}))$. We note that for small systems,  the second assumption is valid, but the first one breaks down because the strength of the system-surrounding interactions is of comparable magnitude to the interactions within the system. In fact, recent work using harmonic oscillators and hard sphere gasses~\citep{dixit2013maximum,dixit2015detecting,dixit2017mini} has shown that the statistics of small systems differs considerably from the Gibbs-Boltzmann distribution; the fluctuations around the mean are much greater than the corresponding predictions. However, the potential failures of the Gibbs-Boltzmann distribution in describing small systems with realistic Hamiltonians, atomic sized magnets for example, are not well understood.

Here, we use the two dimensional Ising model to study the statistics of a small magnet embedded inside a larger magnet. The larger magnet (the `universe') is kept at a constant energy, and the small magnet exchanges energy with its surroundings (the 'bath'). In this setting, we systematically examine the ability of the Gibbs-Boltzmann distribution in capturing the statistics of the small magnet. We find that there are large differences between the predictions from the Gibbs-Boltzmann distribution and the observed statistics of the magnet that depend on both magnet size and the temperature of the surroundings. Specifically, the high energy tails of the distribution are not well captured by the Gibbs-Boltzmann model. Our analysis suggests that the statistical coupling between the system Hamiltonian and the system-bath interaction Hamiltonian predicts whether Gibbs-Boltzmann distribution accurately describes the statistics of the small system. We also find that a superstatistical generalization of the Gibbs-Boltzmann distribution wherein the temperature of the system is allowed to fluctuate fits the statistics better.  Future directions are discussed as well.

\begin{figure}
  \includegraphics[scale = 0.5]{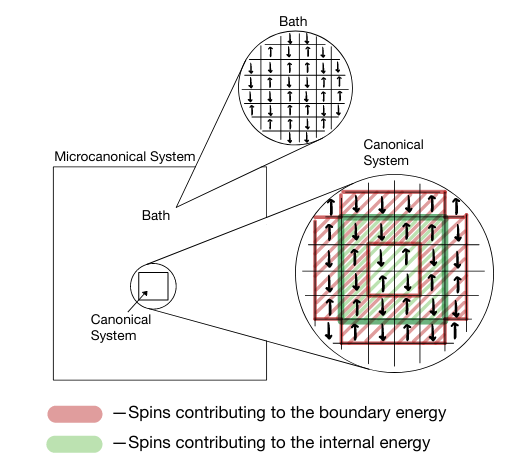}
  \caption{{\bf The schematic of the simulation} The square microcanonical lattice is simulated, and from the interior of that lattice small, canonical spin sites are sampled. It is important to note that while spins that contribute to the boundary energy interact with spins outside of the system, the only {\it interactions} that contribute to the boundary energy are those that are across the boundary.   \label{fg:schem}}
\end{figure}

{\bf Results:} The Hamiltonian of a $N\times N$ two dimensional ferromagnetic Ising model is given by
\begin{eqnarray}
\varepsilon(\bar \sigma) = -J \sum_{\langle i,j \rangle}\sigma_i \sigma_j
\end{eqnarray}
where $\bar \sigma$ is the configurational state of the magnet and $\langle i,j \rangle$ denotes that spins $\sigma_i$ and $\sigma_j$ are nearest neighbors. Individual spins can take on values $\sigma_i = \pm 1$, and the interaction between the spins is mediated by the coupling constant $J > 0$. Without loss of generality, in our calculations we choose $J = 1$. The schematic of our simulation system is shown in Fig.~\ref{fg:schem}, and the simulation procedure is as follows. We construct a large ($100\times 100$) periodic two dimensional Ising lattice which represents the `universe'. The universe is held at a fixed energy $\varepsilon_{\rm tot}$ and evolved using the Demon algorithm~\citep{creutz1983microcanonical}. Thus, the configurations of the universe are sampled from the microcanonical ensemble where each configuration has an equal probability. Within this universe, we focus on small $n\times n$ sites, with $ n\in [3, 10]$, that represent the `system'. The system exchanges energy with its surroundings as the universe samples equal-energy configurations. The energy is exchanged between the system and the bath through the spins at the boundary (see Fig.~\ref{fg:schem}).
This is exactly the setup of the canonical ensemble. Therefore, the statistics of the system should ideally be represented by the Gibbs-Boltzmann distribution.  

Because we want to test the validity of the Gibbs-Boltzmann distribution when applied to small Ising sites, we need to initialize the microcanonical universe at different constant energies. To do this, we first initialize a lattice with spins randomly initialized at each of the lattice points. We evolve the universe according to the canonical ensemble, and after a sufficiently large number of spin flips, we take a snapshot of the universe at temperature $1/\beta$. This snapshot is our initialized microcanonical universe which we will then propagate microcanonically for duration of the sampling procedure. Samples of sites of fixed sizes are randomly sampled at fixed intervals and relevant quantities are recorded: the system configuration, the system energy, and the interaction energy between the system and the surroundings. The details of the simulation procedure and location of scripts are provided in Appendix A.

Using these sampled spin configurations, we fit a Gibbs-Boltzmann model to the distribution of energies by tuning the inverse temperature $\beta$. The distribution is fit by matching the average energy of the system to the one predicted by the Gibbs-Boltzmann distribution. The exact procedure is outlined in more detail in Appendix B. The distribution of energies predicted by the Gibbs-Boltzmann model are of the form
\begin{eqnarray}
p(\varepsilon) \propto \Omega(\varepsilon)\exp \left ( -\beta \varepsilon\right ),
\end{eqnarray}
where $\Omega(\varepsilon)$ is the microcanonical partition function of the small system when held at a constant energy $\varepsilon$.

\begin{figure}
  \includegraphics[scale = 1.25]{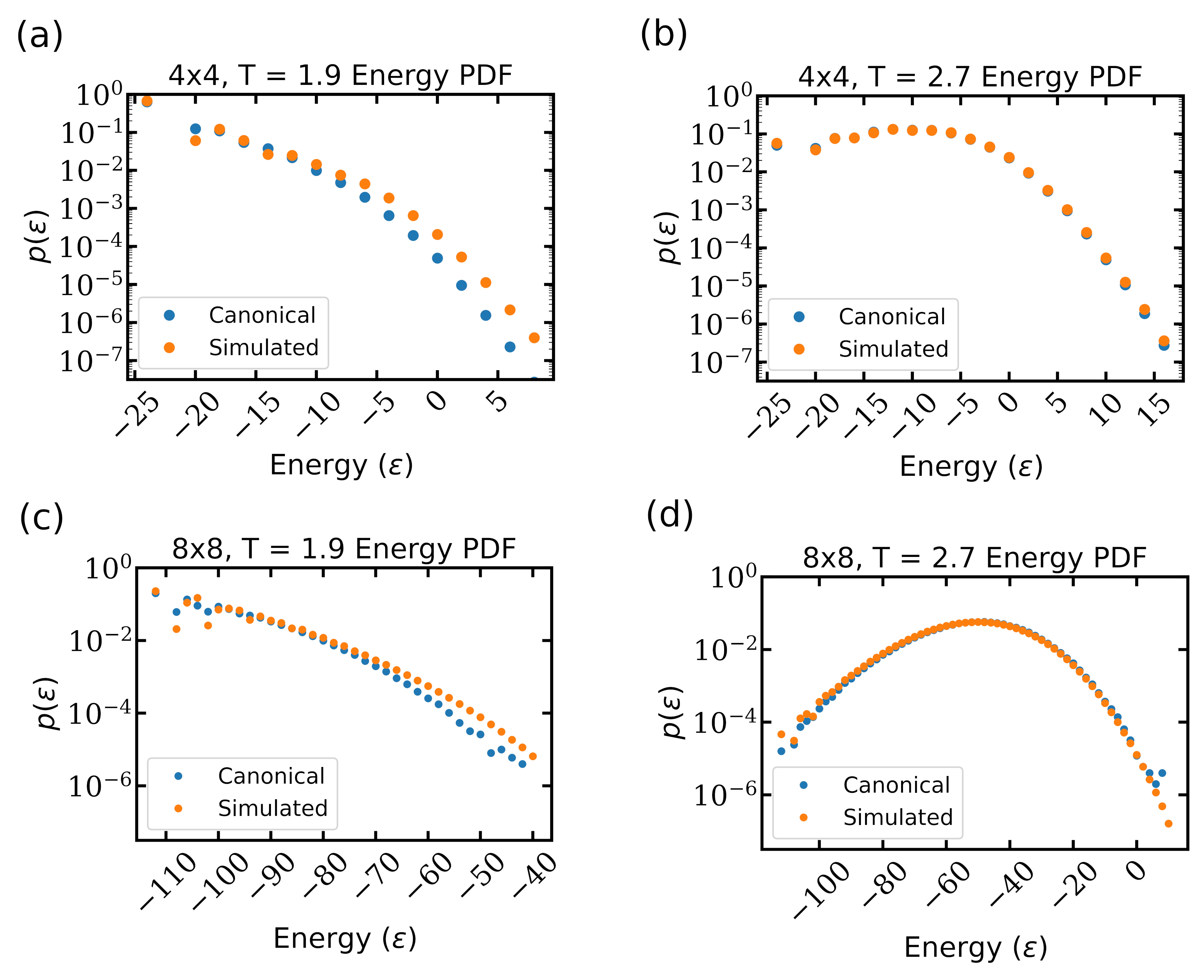}
  \caption{{\bf Distribution of energies in the small system and comparison with the predictions from the Gibbs-Boltzmann distribution.} We have depicted the energy distributions for $4 \times 4$ Ising sites at temperatures (a) $T=1.9$ and (b) $T=2.7$, and for $8 \times 8$ Ising sites at temperatures (c) $T=1.9$ and (d) $T=2.7$. To fit the Gibbs-Boltzmann distribution we tuned $\beta$. We observe deviations among the lowest energies arising from the precise definition of the boundary, and more importantly, we observe systemic deviations between the tails of the Gibbs-Boltzmann distribution and the simulated distribution.  \label{fg:dists}}
\end{figure}
In Fig.~\ref{fg:dists}, we present the distribution of energies $p(\varepsilon)$ as obtained from the simulations and the corresponding Gibbs-Boltzmann fits for a $4\times 4$ and a $8\times 8$ site at temperatures $T = 1.9$ and $T = 2.7$. We see clearly that for temperatures well above the critical point of the infinite two dimensional Ising model ($T \approx 2.27$), the Gibbs-Boltzmann distribution accurately describes the distribution of energies, even for very small systems. At the same time, there are systematic deviations for temperatures below the critical point. First, there is a significant deviation at low energies that arises because of the idiosyncrasies of the definition of the boundary in the Ising model (see Appendix C). Second, and most importantly, the Gibbs-Boltzmann distribution is unable to capture the high energy tail of the distribution.

\begin{figure}
  \includegraphics[scale = 0.75]{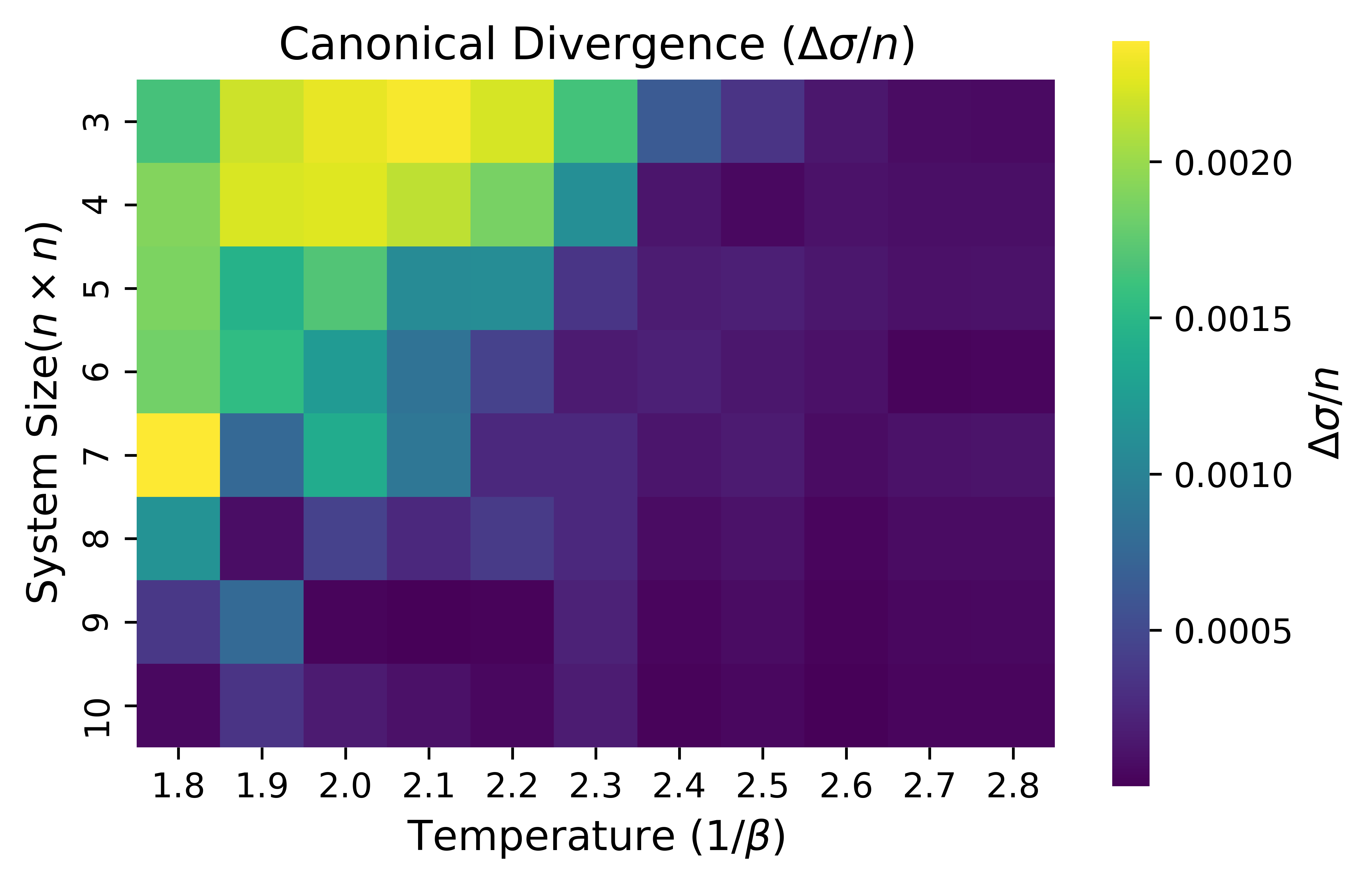}
  \caption{{\bf Accuracy of the Gibbs-Boltzmann distribution in fitting the energy distribution.} The absolute difference in the standard deviation of the simulated energy distribution and the corresponding Gibbs-Boltzmann fit. Because the standard deviation of the energy distributions grows with $n$, the number of spins in any one dimension, and the size of the system grows as $n^2$, the standard deviation differences are normalized by $n$.    \label{fg:heatmap}}
\end{figure}

For a  more insightful analysis, we  fit the Gibbs-Boltzmann distribution to simulated energy distributions for a range of temperatures and systems sizes (Fig.~\ref{fg:heatmap}). Here, we plot the absolute difference in standard deviation of the energy distribution, normalized by $n$, as estimated from the simulation and the Gibbs-Boltzmann fit. The normalization ensures that there are no system-size based artifacts.  We observe two trends. First,  the degree of convergence of the Gibbs-Boltzmann fit depends on both the system size and the temperature of the bath. Notably, as the size of the system increases the deviations between the Gibbs-Boltzmann and simulated distribution occurs at successively lower temperatures, until for an $10 \times 10$ sample we observe the deviations become negligible across all temperatures tested. This size potentially marks the convergence of the system to the canonical ensemble.

\begin{figure}
  \includegraphics[scale = 0.75]{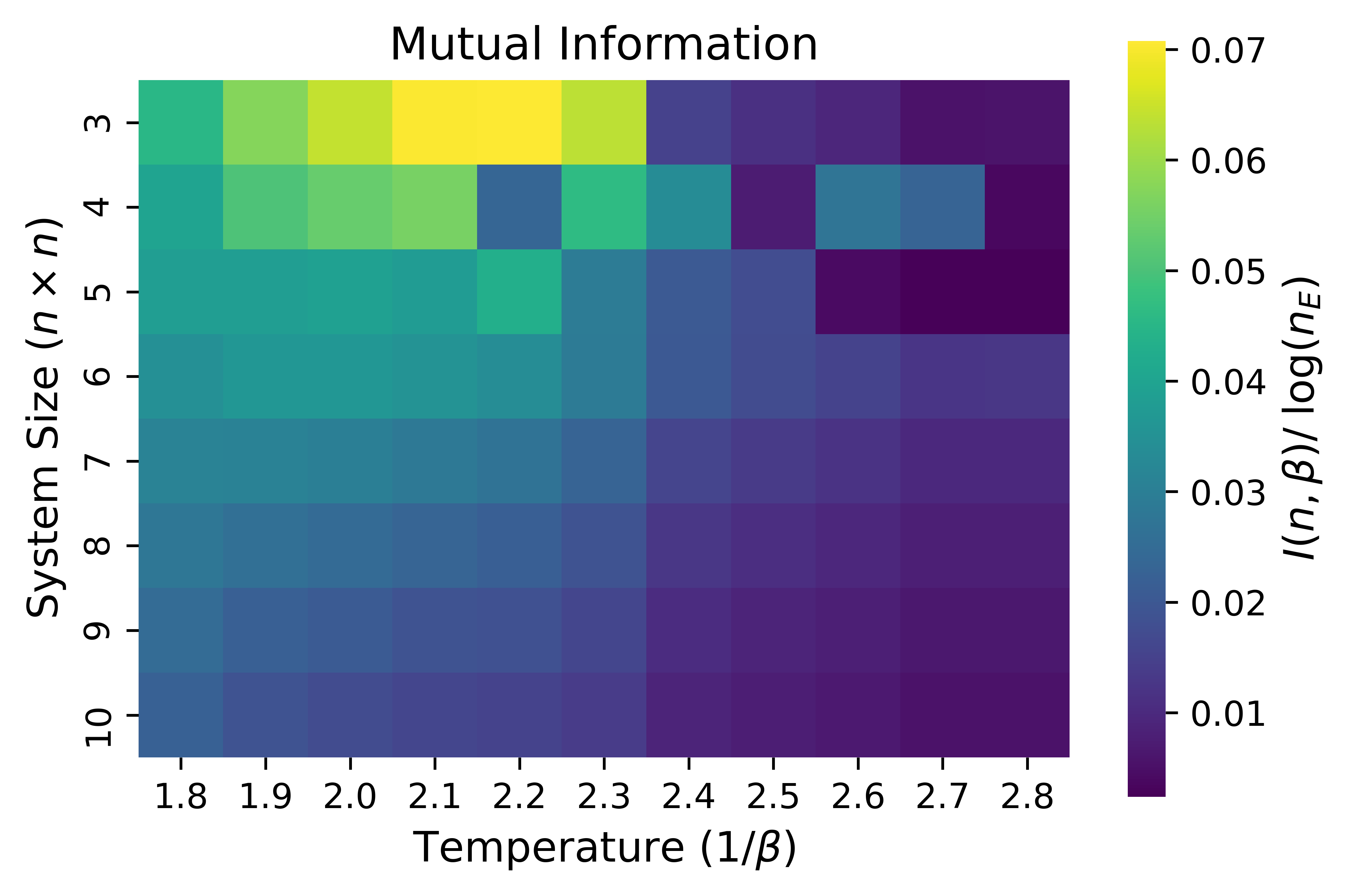}
  \caption{{\bf Mutual information between system energy and the strength of the system-bath boundary interactions.} We have normalized the mutual information by $\text{log} (n_E)$, the number of discrete energies for a $n \times n$ Ising model.  \label{fg:mutual_inf}}
\end{figure}

Additionally, we see the Gibbs-Boltzmann distribution fits the simulated energy  distribution for sufficiently high temperatures, even for the smallest system sizes. The high temperature convergence for all system sizes tested indicates that the assumption of a macroscopic systems may not be necessary in order to apply the Gibbs-Boltzmann model. We hypothesize that the accuracy of the Gibbs-Boltzmann distribution is driven also by the decoupling between the system and the bath. We quantify this decoupling by the mutual information between the internal energy distribution for the system $\varepsilon$, and the system-bath
boundary energy distribution $\varepsilon_{\rm int}$. For the joint distribution $p(\varepsilon, \varepsilon_{\rm int})$ of system size $n$ and temperature $1/\beta$, the mutual
information is defined as
\begin{equation}
  I(n, \beta) = \sum_{\varepsilon, \varepsilon_{\rm int}} p(\varepsilon, \varepsilon_{\rm int})\log  \left(  \frac{p(\varepsilon, \varepsilon_{\rm int})}{p(\varepsilon)p(\varepsilon_{\rm int})} \right).
\end{equation}
The mutual information quantifies the covariance between the two variables $\varepsilon$ and $\varepsilon_{\rm int}$, and is zero {\it iff} the two variables are statistically independent of each other. Fig. \ref{fg:mutual_inf} displays the mutual information between the boundary and internal energy of the system for the range of system sizes and temperatures tested. To avoid system size effects, the mutual information is normalized by the logarithm of the total number of different energy states for the system. The region in which there is a high degree of mutual information between the two distributions coincides with the region where deviations from the Gibbs-Boltzmann distribution are high (see Fig.~\ref{fg:heatmap} and Fig.~\ref{fg:mutual_inf}). This observation suggests that the decoupling between the system and the bath could be a predictor of convergence to the Gibbs-Boltzmann distribution. Furthermore, the Spearman correlation between the mutual information and divergence from the Gibbs-Boltzmann distribution across all temperatures and system sizes is 0.8 ($p < 10^{-20}$). 

{\bf Superstatistics confers a better fit to the statistics of small systems:} How do we fix the discrepancy between the Gibbs-Boltzmann distribution and the observed statistics? One way forward is as follows: from Eq.~\ref{eq:canderive}, we write the probability of observing the system configuration 
\begin{eqnarray}
p({\bf x}) &\propto& \Omega \left ( \varepsilon_{\rm tot} - \varepsilon({\bf x})   - \varepsilon_{\rm int}({\bf x},{\bf y})\right ) \nonumber \\
&=& \exp \left ( S_{\rm bath} \left (  \varepsilon' - \varepsilon({\bf x}) \right )\right )  \nonumber \\
&\approx& \exp \left ( S_{\rm bath} \left (  \varepsilon'  \right ) - \beta' \varepsilon({\bf x})\right )  \propto \exp \left ( -\beta' \varepsilon({\bf x}) \right ).
\end{eqnarray}
Here, we denote by $ \varepsilon' $ sum total of the energy of the bath $\varepsilon_{\rm bath}({\bf y})$ and the interaction energy $\varepsilon_{\rm int}({\bf x},{\bf y})$. For small systems, this term is not negligible. Therefore, the specific value of the Taylor series coefficient $\beta'$ will depend on the specific realization of  the system-bath interaction energy. Moreover, for a given system energy, multiple different values of the system-bath interaction may be permitted. Therefore, the variability in system-bath interactions may be captured by allowing the temperature of the system to vary. This approach is called superstatistics and has been previously used to model the thermodynamics of small systems \citep{dixit2013maximum,dixit2015detecting,dixit2017mini}. Specifically, we have for any configuration of the system ${\bf x}$,
\begin{eqnarray}
p({\bf x}) = \int p({\bf x}|\beta) f(\beta)d\beta
\end{eqnarray}
where $p({\bf x}|\beta)$ is the Gibbs-Boltzmann distribution. The distribution over temperatures is usually taken to be the gamma or inverse gamma distribution. This choice is often justified using the maximum entropy principle \citep{dixit2013maximum} but can also be derived from first principles \citep{Beck2003}. Additionally, it has been demonstrated that for turbulent systems the log-normal distribution may be appropriate \citep{Beck2004}. Here, we choose $f(\beta)$ to be the inverse gamma distribution:
\begin{equation}
  f(\beta ; \alpha, \lambda) = \frac{\lambda^\alpha}{\Gamma(\alpha)}\beta^{-\alpha-1}e^{-\lambda/\beta}.
\end{equation}
\begin{figure*}
    \centering
  \includegraphics[scale=1]{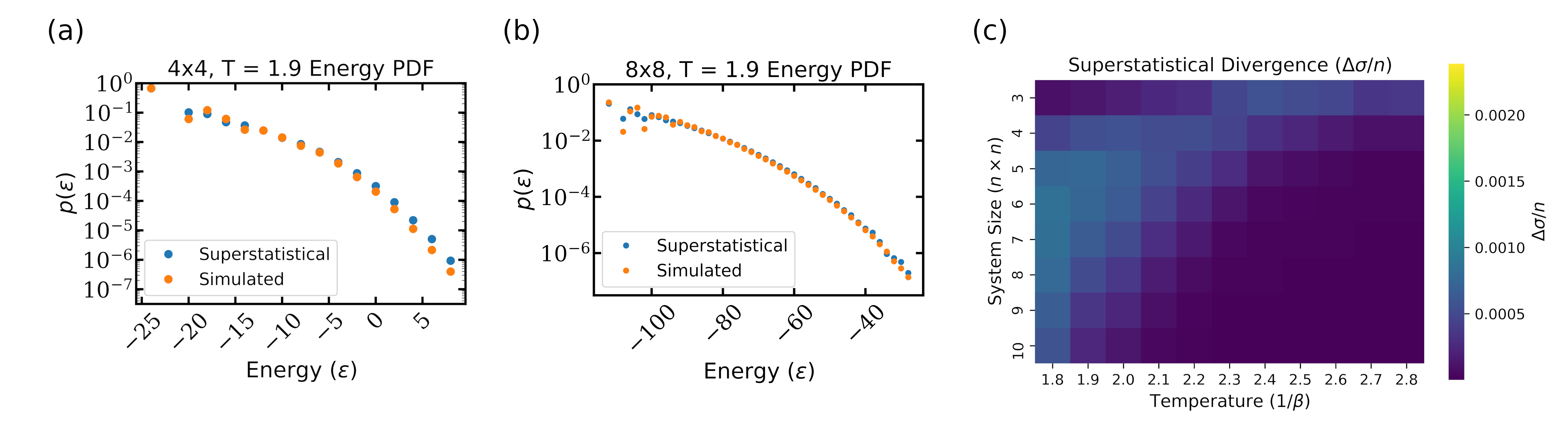}
  \caption{{\bf Superstatistical fits to energy distributions.}  Panels (a) and (b), superstatistical fits to the energy distribution $p(\varepsilon)$ for two system sizes shown in Fig.~\ref{fg:dists}. Panel (c), the heatmap of the absolute difference of normalized standard deviation of the energy distributions from the simulation and the superstatistical fit.\label{fg:superstat}}
\end{figure*}

We fit the parameters $\alpha$ and $\lambda$ of the superstatistical distribution by minimizing the Kullback-Leibler divergence between the energy distributions (see Appendix B). In Fig.~\ref{fg:superstat}, we show the energy distributions observed in the simulation and the corresponding fit from the superstatistical model for the same system sizes in panels (a) and (b) of Fig.~\ref{fg:dists}. Unlike the canonical Gibbs-Boltzmann distribution, the superstatistical distribution tends better predict the high energy tail of the distribution. Moreover, the absolute difference in the variance in energy  between the simulated data and the superstatistical model is smaller for all system sizes, as shown in panel (c) of Fig.~\ref{fg:superstat}. The heatmap shows that the superstatistical fit is uniformly better than the Gibbs-Boltzmann distribution. Is this surprising? We note that because the Dirac Delta function $\delta(\beta)$ is a special case of the inverse gamma distribution, the Gibbs-Boltzmann distribution is a sub-family of the superstatistical distribution. Therefore, the superstatistical approach is guaranteed to be at least as accurate as the canonical ensemble. Yet, given that these two are nested models, we can statistically evaluate whether it is justifiable to include additional complexity to the model. We use the likelihood ratio test to test whether the fit given by the superstatistical approach is statistically significant.  Indeed, we find that for all tested distributions, the chi-squared test derived from the likelihood ratio indicates that the superstatistical model fits the data better in a statistically significant manner ($\chi^2 > 6000, p < 10^{-20}$).

Importantly, the superstatistical approach allows us to not only fit the simulated distributions better, but also provides a quantitative characterization of the departure from Gibbs-Boltzmann distribution. As mentioned above, the Gibbs-Boltzmann distribution with a unique temperature is a subset of the superstatistical distribution with the temperatures distributed according to a Dirac Delta function. Therefore, we can use the width (coefficient of variation) of the $p(\beta)$ distribution to measure the departure from canonical behavior. Indeed, we find that the coefficient of variation, defined as $\sigma / \mu$ of $p(\beta)$, is significantly correlated with the error in the Gibbs-Boltzmann fit (the absolute difference of normalized standard deviations, see Fig.~\ref{fg:heatmap}) with a Spearman correlation of 0.75 ($p \sim 10^{-17}$).

{\bf Discussions:} We studied the performance of the Gibbs-Boltzmann distribution in modeling thermal statistics of small Ising models. We showed that systematic deviations exist that   depend on both the size and temperature of the system. We also showed that a superstatistical approach that allows the temperature of the system to fluctuate was a superior descriptor of its statistics. Notably, the superstatistical approach has the Gibbs-Boltzmann statistics as a special case and can thus be viewed as a generalization of the canonical ensemble that is useful to model systems that exchange energy with their surroundings, regardless of their size.  However, this is not to say that the superstatistical model is perfectly able to model the data. In the cases where the canonical distribution fails to model the simulated data, the superstatistical model also fails to model the simulated data, but to a far lesser extent. 

Regardless, generalizing the Gibbs-Boltzmann distribution by superstatistics has been shown to result in a increase in our ability to model the energy distribution of a small Ising model, beyond the mere addition of  degrees of freedom. Taking the superstatistical approach to modeling small systems requires further consideration. In particular, investigating the ability of the superstatistical distribution to model other types of small systems, both strongly and weakly interacting, would provide a clearer picture of the limitations of the superstatistical approach. Additionally, investigating the precise ways in which the canonical model fails may lead to a more concrete way of determining when the Gibbs-Boltzmann distribution may be applied. We have shown that for the Ising model the degree of deviation is dependent of both the temperature and size of the system. Furthermore, we have proposed examining the mutual information between the boundary and internal energy of the system to gauge the degree of decoupling between the system and its environment, and consequently the degree of deviation from the canonical distribution.


\newpage

\appendix
\section{Simulation Procedure}
The procedure used for simulating a microcanonical universe is as follows. We initialize a $100 \times 100$ lattice with spins randomly initialized with values $ + 1$ or $- 1$, corresponding to spin up and spin down respectively. This lattice will have a high temperature since the spins are randomly initialized. We then anneal the lattice to a desired temperature using the Metropolis-Hastings algorithm \citep{Hastings1970}. Once the spin dynamics of the lattice reaches a steady state, we use a snapshot of the $100 \times 100$ lattice, with energy $E_0$ as the starting point of the microcanonical simulation. Since there is no strict definition of time for a Monte-Carlo simulation, we have adopted the convention that each time a spin flip is proposed, and the accepted or rejected, one time step progresses. In our simulation, the lattice was evolved using the Metropolis-Hastings algorithm over $10^5$ time steps.

The lattice is evolved microcanonically according to the Demon algorithm \citep{creutz1983microcanonical}, which allows for efficient simulation at an approximately constant energy. This is achieved by the introduction of an energy "bank" that acts as a bookkeeping device. We denote the energy contained in the bank as  $E_{\mathbf B} \in [0, E_{\mathbf M}]$. A brief outline of the algorithm is as follows. A spin is randomly selected from the lattice, and a spin flip is proposed. If the spin flip requires energy $E'$, the flip is only accepted if $E_{\mathbf B} \geq E'$, and in accepting the spin flip we deduct $E'$ from the bank. If the proposed spin flip decreases the energy of the lattice, we accept the spin flip only if $E_{\mathbf B} + E' \leq E_{\mathbf M}$, and when the spin flip is accepted $E'$ is added to the bank. Using this algorithm the energy of the microcanonical lattice $E$ is restricted to $E \in [E_0, E_{\mathbf M}]$.  Importantly, the results of the sampling procedure are robust with respect to choice of $E_{\mathbf{M}}$ so long as $E_{\mathbf{M}} \ll E_0$.

As the microcanonical universe is evolved, we randomly sample small $n \times n$ sites from the interior of the lattice. The sampling frequency is determined by the decorrelation time of the energy of successive samples. For each sample we record the internal energy of the $ n \times n$ site, and the interaction energy across the system-bath boundary. In our simulations a sampling frequency of 100 spin flips was used over a total simulation length of $10^9$ time steps. Additionally, where possible, we record the particular configuration of the sample as an integer.
\par
The simulation scripts can be found at \url{https://github.com/lukasherron/ising-model}.

\section{Data Analysis Procedure}
In an attempt to mitigate the effect of undersampling, we restrict the simulated distributions to energies $\varepsilon$ such that $p(\varepsilon) > 10^-7$. Then, to fit the Gibbs-Boltzmann distribution to the simulated energy distribution, we tune the inverse temperature $\beta$ so that the average energies of the two distributions are the same.
The exact procedure is different for sites smaller and larger than $4 \times 4$. For sites of size $4 \times 4$ and below, we are able to enumerate each microstate of the site, and are thus able to directly compute the Gibbs-Boltzmann distribution. For sites larger than $4 \times 4$, there are more microstates than we are able to enumerate, so we fit the simulated distribution by simulating a canonical system over a range of temperatures, and finding the temperature of the canonical simulation that matches the average energy of the distribution generated during our main simulation procedure. The parallel simulations used for fitting the Gibbs-Boltzmann distribution are also used in fitting the superstatistical model, and are run over $5 \times 10^8$ time steps.
\par
To fit the superstatistical distribution, we tested the free parameters $\lambda$ and $\alpha$ of the inverse gamma distribution for $\beta$ through a factorial design - that is to say we test every $(\lambda, \alpha)$ pair in the parameter space. The optimal values $\lambda$ and $\alpha$ are found to take on vary over up to three orders of magnitude, so for computational efficiency we test equally spaced $(\lambda, \alpha)$ pairs in the parameter log-space. Since $\lambda$ and $\alpha$ are found to scale similarly this does not negatively impact the accuracy of the fitting procedure. We then minimize the Kullback-Leibler divergence to fit the superstatistical model. The KL divergence is defined as
\begin{eqnarray}
D(P_{\text{sim}}(\varepsilon) || P_{\text{model}}(\varepsilon)) = \sum_{\varepsilon} P_{\text{sim}}(\varepsilon) \; \text{log} \left( \frac{P_{\text{sim}}(\varepsilon)}{P_{\text{model}}(\varepsilon)} \right),
\end{eqnarray}
where $P_{\text{sim}}(\varepsilon)$ and $P_{\text{model}}(\varepsilon)$ are the probabilities of observing energy $\varepsilon$ in the simulated distribution and superstatistical model respectively. The KL divergence was specifically chosen because minimizing $D(P_{\text{sim}}(\varepsilon) || P_{\text{model}}(\varepsilon))$ is equivalent to maximizing the data likelihood of $P_{\text{sim}}(\varepsilon)$ given $P_{\text{model}}(\varepsilon)$. This ensures that when we are using the likelihood ratio test to see if our results are significant we are using the maximum likelihood superstatistical model.
\par
The data analysis script can be found at \url{https://github.com/lukasherron/ising-model}.

\section{Low Energy Deviations}
The low energy deviations that are present for all site size and temperatures tested arises due to the way in which the boundary is defined for an embedded canonical system. Consider two identical canonical systems $A$ and $B$, except system $A$ is {\it not} embedded in bath, and system $B$ is embedded in a bath. The evolution of both systems may be described by the Metropolis-Hastings algorithm. Note that the lowest energies occur when site is magnetized - most of the spins face the same direction. In particular, the lowest energy state occurs when the site is completely magnetized. The next few lowest energy states occur when spins along the boundary are flipped. To illustrate why the probabilities of the lowest energy states differ between the simulated and Gibbs-Boltzmann distribution, consider the probability that a corner spin is flipped in systems $A$ and $B$. More specifically, consider the situation where $A$ and $B$ are at the same temperature and completely magnetized, and the corner spin is flipped so that the energy of the system increases.

According to the Metropolis-Hastings algorithm, when a proposed spin flip increases the energy of the system by $\Delta E$, the spin flip is accepted only if a uniformly distributed random variable $x \in [0,1]$ satisfies $ x < \exp ({-\beta \Delta E})$. Let us examine the average positive $\Delta E$, denoted as $\langle \Delta E^+ \rangle $, for a corner spin in systems $A$ and $B$. For system $A$, which is not embedded in a bath, the corner spin has two nearest neighbors so that $\langle \Delta E^+ \rangle = 4J$. And for embedded system $B$, the corner spin has four nearest neighbors so that $\langle \Delta E^+ \rangle = 4.8J$. Thus, we can expect that a corner spin in an embedded canonical system will be flipped more often than a corner spin in an isolated canonical system, which the Gibbs-Boltzmann distribution describes. This is the only way that the second lowest energy state in a square Ising model can occur, and is exactly what we observe in our analysis (see Fig. 2). A similar analysis may be applied to edge spins and other energy levels, but the number of microstates corresponding to each energy level of the Ising model grows rapidly.

These low energy deviations are not due to an error due to simulation or analysis, but rather due to the precise way in which the boundary is defined. It may be of interest to determine if there is an optimal way to define the boundary such that the low-energy deviations are minimal.

\end{document}